\documentclass[prd,superscriptaddress,amsfonts,amssymb,amsmath,showpacs,onecolumn]{revtex4-2}

\usepackage{bm}
\usepackage{amsfonts}
\usepackage{latexsym}
\usepackage[latin1]{inputenc}
\usepackage{graphicx}
\usepackage{amsmath}
\usepackage{palatino}
\usepackage{mathpazo}
\usepackage{textcomp}
\usepackage{float}
\usepackage{booktabs}
\usepackage{dcolumn}
\usepackage{ragged2e}
\usepackage{hyperref}
\hypersetup{colorlinks,citecolor=blue}
\hypersetup{colorlinks=true,linkcolor=red,filecolor=magenta,    urlcolor=blue}
\usepackage{amsmath}
\usepackage{graphicx}
\usepackage{makecell}
\usepackage{hyperref}
\usepackage[mathlines]{lineno}
\usepackage{amssymb}
\usepackage{enumitem}
\usepackage{mathtools}
\usepackage{mathrsfs}
\usepackage{setspace}
\usepackage{color}
\usepackage{graphicx}
\usepackage[dvipsnames]{xcolor}
\usepackage{mathalfa}
\usepackage{calligra}
\usepackage[misc]{ifsym}
\DeclareMathAlphabet{\mathpzc}{OT1}{pzc}{m}{it}
\DeclareMathAlphabet{\mathcalligra}{T1}{calligra}{m}{n}
\hypersetup{colorlinks=true,citecolor=blue,linkcolor=red,urlcolor=magenta}
\usepackage{xcolor}
\usepackage{orcidlink}
\usepackage[caption=false]{subfig}
\usepackage{commath}
\def\jnl@style{}
\def\aaref@jnl#1{{\jnl@style#1}}

\def\aaref@jnl#1{{\jnl@style#1}}

\def\aj{\aaref@jnl{AJ}}                   
\def\apj{\aaref@jnl{ApJ}}                 
\def\apjl{\aaref@jnl{ApJ}}                
\def\apjs{\aaref@jnl{ApJS}}               
\def\apss{\aaref@jnl{Ap\&SS}}             
\def\aap{\aaref@jnl{A\&A}}                
\def\aapr{\aaref@jnl{A\&A~Rev.}}          
\def\aaps{\aaref@jnl{A\&AS}}              
\def\mnras{\aaref@jnl{Mon.~Not.~Roy.~Astron.~Soc.}}             
\def\prd{\aaref@jnl{Phys.~Rev.~D}}        
\def\plb{\aaref@jnl{Phys.~Lett.~B}}        
\def\prc{\aaref@jnl{Phys.~Rev.~C}}  
\def\prl{\aaref@jnl{Phys.~Rev.~Lett.}}    
\def\qjras{\aaref@jnl{QJRAS}}             
\def\skytel{\aaref@jnl{S\&T}}             
\def\ssr{\aaref@jnl{Space~Sci.~Rev.}}     
\def\zap{\aaref@jnl{ZAp}}                 
\def\nat{\aaref@jnl{Nature}}              
\def\aplett{\aaref@jnl{Astrophys.~Lett.}} 
\def\apspr{\aaref@jnl{Astrophys.~Space~Phys.~Res.}} 
\def\physrep{\aaref@jnl{Phys.~Rep.}}      
\def\physscr{\aaref@jnl{Phys.~Scr}}       
\def\commat{\aaref@jnl{Comm.~Math.~Phys.}}              
\def\science{\aaref@jnl{Science}}               
\def\cqg{\aaref@jnl{Classical Quant.~Grav.}}            
\def\jpcs{\aaref@jnl{JPCS}}                                     
\def\ijmpd{\aaref@jnl{Int.~J.~Mod.~Phys.~D}}                    
\def\grg{\aaref@jnl{Gen.~Relat.~Gravit.}}               
\def\rpp{\aaref@jnl{Rep.~Prog.~Phys.}}          
\def\npa{\aaref@jnl{Nucl.~Phys.~A}}        
\def\lrr{\aaref@jnl{Living Rev.~Rel.}}                   
\def\jcap{\aaref@jnl{J.~Cosmology Astropart.~Phys.}}    
\def\rmp{\aaref@jnl{Rev.~Mod.~Phys.}}   
\def\epjc{\aaref@jnl{Eur.~Phys.~J.~C}}

\allowdisplaybreaks[1]
\renewcommand{\arraystretch}{1.1}
\begin{document}

\preprint{APS/123-QED}

\title{On possible wormhole solutions supported by non-commutative geometry within $\mathpzc{f}(\mathcal{R},\mathscr{L}_m)$ gravity}

\author{N. S. Kavya\orcidlink{0000-0001-8561-130X}}
\email{kavya.samak.10@gmail.com}
\affiliation{Department of P.G. Studies and Research in Mathematics,
 \\
 Kuvempu University, Shankaraghatta, Shivamogga 577451, Karnataka, INDIA
}%

\author{V. Venkatesha\orcidlink{0000-0002-2799-2535}}%
 \email{vensmath@gmail.com}
\affiliation{Department of P.G. Studies and Research in Mathematics,
 \\
 Kuvempu University, Shankaraghatta, Shivamogga 577451, Karnataka, INDIA
}%


\author{G. Mustafa\orcidlink{0000-0003-1409-2009}}%
\email{gmustafa3828@gmail.com}
\affiliation{
 Department of Physics, Zhejiang Normal University, Jinhua, 321004, People's Republic of China.
}%

\author{P.K. Sahoo\orcidlink{0000-0003-2130-8832}}
\email{pksahoo@hyderabad.bits-pilani.ac.in}
\affiliation{
 Department of Mathematics, Birla Institute of Technology and Science-Pilani,\\
 Hyderabad Campus, Hyderabad 500078, INDIA
}%


\date{\today}

\begin{abstract}
Non-commutativity is a key feature of spacetime geometry. The current article explores the traversable wormhole solutions in the framework of $\mathpzc{f}(\mathcal{R},\mathscr{L}_m)$ gravity within non-commutative geometry. By using the Gaussian and Lorentzian distributions, we construct tideless wormholes for the nonlinear $\mathpzc{f}(\mathcal{R},\mathscr{L}_m)$ model $\mathpzc{f}(\mathcal{R},\mathscr{L}_m)=\dfrac{\mathcal{R}}{2}+\mathscr{L}_m^\alpha$. For both cases, we derive shape functions and discuss the required different properties with satisfying behavior. For the required wormhole properties, we develop some new constraints. The influence of the involved model parameter on energy conditions is analyzed graphically which provides a discussion about the nature of exotic matter. Further, we check the physical behavior regarding the stability of wormhole solutions through the TOV equation. An interesting feature regarding the stability of the obtained solutions via the speed of sound parameters within the scope of average pressure is discussed. Finally, we conclude our results.
\begin{description}
\item[Keywords]
Traversable wormhole, $\mathpzc{f}(\mathcal{R},\mathscr{L}_m)$ gravity, energy conditions, non-commutative geometry,\\ equilibrium condition.

\end{description}
\end{abstract}

\maketitle

\section{INTRODUCTION}\label{I}
    \par Wormholes are tube-like structures whose mouths at both ends connect distinct regions positioned in the same universe or different universes. Initially, Flamm proposed the existence of these hypothetical connections in the universe \cite{flamm}. With the isometric embedding, he probed Schwarzschild solutions to governing gravity relations. Einstein and Rosen gave the mathematical vision of a bridge-like structure having an event horizon \cite{erbridge}. A traversable wormhole solution (horizon-less scenario) to the classical Einstein field equation was put forth by Morris and Thorne \cite{morrisandthorne}. These wormholes violate the energy conditions, specifically the Null Energy Condition (NEC). Therefore, to accomplish the traversability of wormhole in the realm of GR, a certain type of hypothetical fluid disobeying NEC is required (as an ordinary matter agrees with the known laws of physics). In the framework of GR, the wormhole facilitated with non-hypothetical fluid cannot be formulated. Accordingly, numerous endeavors took place to reduce the usage of exotic matter \cite{em1,em2,em3,em4,em5,em6,em7,em8}. Modified theories, on the other hand, gave a satisfactory solution to the exotic matter problem. B\"{o}hmer et al. \cite{ec1} constructed wormhole structures with specific redshift and shape functions in modified teleparallel gravity that obeys energy conditions. In the background of $\mathpzc{f}(\mathcal{R})$ gravity, Lobo and Oliveira \cite{ec2} studied traversable wormhole geometries. They imposed that matter threading the wormhole satisfies energy conditions. In \cite{ec3}, authors formulated wormholes without exotic matter in the Einstein-Gauss-Bonnet theory. Capozziello et al. \cite{CZ1,CZ2,CZ3,CZ4,CZ5,CZ6,CZ7,CZ8,Luongo3} explored various aspects of WH in alternative gravity theories. Recently, Capozziello and Nisha \cite{CZ9} considered non-local gravity in view of obtaining stable and traversable wormhole solutions.

    \par In recent years, the investigation of wormhole geometry has piqued the interest of many astrophysicists. Rahaman et al. \cite{ref1}, presented a solution for a wormhole with phantom energy in spherically symmetric spacetime. Their model suggests the presence of a wormhole supported by an arbitrarily small amount of phantom energy. Dynamic thin shell traversable wormhole is examined based on the black-bounce spacetimes \cite{ref2}. In \cite{ref3}, Zubair et al. investigated static spherically symmetric wormhole geometry with anisotropic, isotropic, and barotropic matter content. A four-dimensional wormhole solution with 'Casimir-like' energy is put forth by Maldacena and the team \cite{ref4}, which in ambient space doesn't lead to causality violation. Gravitational lensing is studied by rotating and non-rotating Damour-Solodukhin wormholes using the Gauss-Bonnet theorem and Bozza's method \cite{ref5}. The effects of repulsive gravity in gravitational contexts have been studied in \cite{Luongo1,Luongo2}. Numerous endeavors have taken place to probe wormhole spacetime in the context of metric affine theories \cite{mat,mat1}, geometry-matter couplings \cite{gmc1,gmc2,gmc3,gmc4}, braneworld \cite{bw}, and non-commutative geometry \cite{ncg,ncg1}.
    
    \par In exploring the manifolds' features in different lights, the concept of non-commutative geometry is phenomenal. In the analysis of spacetime structure, non-commutativity can be presented via modified matter sources. Primarily, it unifies weak and strong forces with gravitational force. P. Aschieri et al \cite{nc1} have constructed classical governing equations of gravity on non-commutative geometry.  Both the deformation of the spacetime geometry and the quantization can be effectively dealt with non-commutativity. In D-brane \cite{nc2}, the coordinates of spacetime can be treated as non-commutative operators represented by $[y^a,y^b]=\mathit{i}\theta^{ab}$. Such operators lead to spacetime discretization. This is indicated by the second-order antisymmetric matrix $\theta^{ab}$ \cite{nc3,nc4,nc5,nc6}. One can remark that non-commutativity substitutes smear objects for point-like structures. The smearing phenomenon can be employed by replacing the Dirac delta function with a Gaussian and a Lorentzian distribution of minimal length $\sqrt{\theta}$. In \cite{nc7}, Schneider et al. have discussed the Gaussian and Lorentzian distributions via simple smearing of a matter distribution within the black hole. With Gaussian distribution, Sushkov examined the wormholes supported by phantom energy \cite{nc8}. Kuhfittig, in \cite{nc9}, shows that certain thin-shell wormholes that are unstable in GR behave stable as a consequence of non-commutativity. The physical impact of the short separation of non-commutative coordinates can be seen in \cite{nc10}. Rahaman et al. investigated wormhole geometry with Gaussian distribution and proved the feasibility of solution in four and five dimensions \cite{nc11}.  
    
    \par For the static spherically symmetric point-like gravitational source having total mass $M$, Gaussian and Lorentzian distribution of energy densities are given by \cite{intrinsic,nc},
    \begin{eqnarray}
        \label{gauss}\rho= \frac{M e^{-\frac{r^2}{4 \theta }}}{8 \pi ^{\frac{3}{2}} \theta ^{\frac{3}{2}}},\\
        \label{lorentz}\rho=\frac{\sqrt{\theta } M}{\pi ^2 \left(\theta +r^2\right)^2}.
    \end{eqnarray}
     
     These choices reflect the notion that the source is spread out or smeared rather than being concentrated at a single point. This is mainly because of the intrinsic uncertainty in the coordinate commutator. Further, the noncommutative correction becomes significant in a region near the origin, specifically when $r \lesssim \theta$. Within this neighborhood, the effects of noncommutativity regularize both the radial and tangential pressures, as well as the matter density. From the particular choice \eqref{gauss} and \eqref{lorentz}, the physical parameters (especially energy density) are finite and asymptotically vanish, supporting the vacuum solution at points far away from the origin. 
    \par In the present manuscript, we attempt to study static spherical symmetric Morris-Thorne wormhole structure in the paradigm of the newly proposed $\mathpzc{f}(\mathcal{R},\mathscr{L}_m)$ gravity. This article is organized as follows: Section \ref{II} provides a mathematical outline of the modified theory in which we discuss governing equations, wormhole solutions in  $\mathpzc{f}(\mathcal{R},\mathscr{L}_m)$ gravity, and energy conditions. In section \ref{III}, we examine the wormhole model with Gaussian and Lorentzian distribution and derive the corresponding shape functions. Also, we analyze the influence of model parameter on the shape functions and energy conditions. Section \ref{IV} assesses the stability of wormholes using the TOV equation. In section \ref{V} we interpret the physical aspects of the wormhole by examining  average pressure and speed of sound. Finally, section \ref{VI} gives the discussion of results and concluding remarks.
	
\section{Mathematical Formulation of the modified theory}\label{II}
\subsection{Initiation}

\par The observational constraints have drawn consideration to some of the shortcomings of GR, which affect it on both smaller and larger scales, such as quantum scale and galactic systems, where modifications to the standard action are required to maintain GR as the fundamental theory of gravity. In other words, it is plausible that the gravitational aspect of the standard GR model needs further examination to address these observable issues. This notion could come in the form of generalizations beyond GR that could serve as an alternative to the formulation. For instance, $\mathpzc{f}(\mathcal{R})$ gravity is one of the most prominent modified geometry theories whose models can agree with observational data. It can reasonably explain the late-time acceleration \cite{fr1,Luongo4} and cosmic inflation \cite{fr2} due to the replacement of the Ricci scalar with its arbitrary function.  A generalized gravity version of  $\mathpzc{f}(\mathcal{R})$ theory is presented in \cite{frlm}. Here, along with the modified geometry section, an explicit form of matter source is coupled to achieve extension to the matter sector of the standard model of particle physics. The modified action for the theory is described as,	
\begin{equation}\label{action}
			S=\int \mathpzc{f}(\mathcal{R},\mathscr{L}_m)	\sqrt{-g}\, d^4x,
		\end{equation}	
where, $\mathpzc{f}$ represents an arbitrary function of scalar curvature $\mathcal{R}$ and the matter Lagrangian $\mathscr{L}_m$. For $\mathpzc{f}=\mathcal{R}/2+\mathscr{L}_m$, one can retain the governing equations of GR. The explicit coupling between geometry and the matter sector results in obtaining a non-vanishing covariant derivative of the Energy-Momentum Tensor (EMT) i.e. $\nabla_a \mathcal{T}^{ab}\ne0$. Due to this, the motion of test particles takes a non-geodesic path that influences the violation of the equivalence principle. Various forms of $\mathscr{L}_m$, representing matter sources, lead to extra force orthogonal to four-velocity \cite{lmrho2,extraforce,lmp1,lmp2}. Recent studies suggest that this theory can be regarded as a possible explanation for cosmic acceleration and dark energy \cite{frlm1,frlm2,frlm3,frlm4,frlm5}. The primary purpose of this manuscript is to assess the wormhole geometry with non-commutativity. In the next section, we shall discuss the governing equation of $\mathpzc{f}(\mathcal{R},\mathscr{L}_m)$ gravity. 
  
\subsection{Governing Equations in $\mathpzc{f}(\mathcal{R},\mathscr{L}_m)$ Gravity}
	 Action describes the governing equation of a gravity theory. With the help of \eqref{action} we can derive the field equations of $\mathpzc{f}(\mathcal{R},\mathscr{L}_m)$ gravity. By varying \eqref{action} with respect to $g^{ab}$, the field equation is obtained as,  
		\begin{equation}\label{fieldequation1}
			\begin{split}
				\mathpzc{f}_\mathcal{R}\mathcal{R}_{ab}+(g_{ab}\nabla_a\nabla^{a}-\nabla_a\nabla_b)\mathpzc{f}_\mathcal{R}-\dfrac{1}{2}\left[\mathpzc{f}- \mathpzc{f}_{\mathscr{L}_m}\mathscr{L}_m\right]g_{ab}=\dfrac{1}{2}\mathpzc{f}_{\mathscr{L}_m}\mathcal{T}_{ab}.
			\end{split}
		\end{equation}
		Here, $\mathpzc{f}_{\mathscr{L}_m}$ and $\mathpzc{f}_\mathcal{R}$ represents the partial derivative of $\mathpzc{f}$ with respect to matter Lagrangian $\mathscr{L}_m$ and the Ricci scalar $\mathcal{R}$ respectively. The Energy-Momentum tensor (EMT) $\mathcal{T}_{ab}$ is defined as,
		\begin{equation}\label{emt}
			\mathcal{T}_{ab}=-\dfrac{2}{\sqrt{-g}} \dfrac{\delta(\sqrt{-g}\mathscr{L}_m)}{\delta g^{ab}}=g_{ab}\mathscr{L}_m-2\dfrac{\partial \mathscr{L}_m}{\partial g^{ab}}.
		\end{equation}  
		By taking the covariant divergence of EMT we get,
		\begin{equation}\label{divofT}
			\nabla^a \mathcal{T}_{ab}=2\left\lbrace \nabla^a \text{ln}\left[\mathpzc{f}_{\mathscr{L}_m} \right]\right\rbrace \dfrac{\partial \mathscr{L}_m }{\partial g^{ab}}. 
		\end{equation}
		\par Now contracting the governing equation \eqref{fieldequation1} we obtain the following correspondence between matter Lagrangian and the trace of EMT:
		\begin{equation}\label{traceoffieldequation}
			\begin{split}
				3\nabla_a\nabla^{a}\mathpzc{f}_\mathcal{R}+\mathpzc{f}_\mathcal{R}\mathcal{R}-2\left[\mathpzc{f} -\mathpzc{f}_{\mathscr{L}_m}\mathscr{L}_m\right]=\dfrac{1}{2}\mathpzc{f}_{\mathscr{L}_m}\mathcal{T}.
			\end{split}
		\end{equation}
		Using the above equation, one can get another form of the field equation,  
		\begin{equation}\label{fieldquation2}
			\begin{split}
				\mathpzc{f}_\mathcal{R}\left( \mathcal{R}_{ab}-\dfrac{1}{3}\mathcal{R}g_{ab}\right) + \dfrac{g_{ab}}{6}\left[\mathpzc{f} -\mathpzc{f}_{\mathscr{L}_m}\mathscr{L}_m\right]=\dfrac{1}{2}\left(\mathcal{T}_{ab} -\dfrac{1}{3}\mathcal{T}g_{ab}\right)\mathpzc{f}_{\mathscr{L}_m}(\mathcal{R},\mathscr{L}_m)+\nabla_a\nabla_{b}\mathpzc{f}_\mathcal{R}.
			\end{split}
		\end{equation}	

        The effective EMT is given by
        \begin{equation}
             \mathcal{T}_{ab}^{eff}=\frac{1}{\mathpzc{f}_{\mathcal{R}}}\left[\dfrac{1}{2}\left(\mathpzc{f}- \mathcal{R}\mathpzc{f}_{\mathcal{R}}\right)g_{ab}-(g_{ab}\nabla_a\nabla^{a}-\nabla_a\nabla_b)\mathpzc{f}_\mathcal{R}+\frac{1}{2}\mathpzc{f}_{\mathscr{L}_m}\mathscr{L}_m g_{ab}+\frac{1}{2}\mathpzc{f}_{\mathscr{L}_m}\mathcal{T}_{ab}\right].
        \end{equation}
    
        The EMT \eqref{emt} for anisotropic matter becomes,
        \begin{equation}\label{energymomentumtensor}
			\mathcal{T}_{ab}=(\rho+p_\tau)\eta_a \eta_b-p_\tau\,g_{ab}+(p_r-p_\tau)\xi_{a}\xi_b,
        \end{equation}
        where, the 4-velocities $\eta^a$ and $\xi^a$ satisfies $\eta^a\eta_a=-1=-\xi^a\xi_a.$
\subsection{Wormhole Solution in $\mathpzc{f}(\mathcal{R},\mathscr{L}_m)$ Gravity}
   The Morris-Thorne metric for the traversable wormhole is described as,
 \begin{equation}\label{whmetric}
		ds^2=e^{2R_f(r)}dt^2-\dfrac{dr^2}{1-\dfrac{S_f(r)}{r} }  - r^2\left(d\theta^2+\text{sin}^2\theta \,d\phi^2\right), 
	\end{equation}
    where $R_f(r)$ and $S_f(r)$ are respectively redshift and shape functions. The redshift function $R_f$ takes a finite value in the entire spacetime to avoid the presence of a horizon. Here in our study, to reduce the complexity of the problem we take $R_f$ as a constant i.e., we are investigating wormhole in the zero tidal force scenario \cite{tidal}. The radial coordinate r takes the values ranging from $r_0$ to $\infty$. The minimum value $r_0$ is called the throat radius and is the fixed point of the shape function $S_f(r)$ i.e., $S_f(r_0)=r_0$. The shape function is significant in achieving the traversability of a wormhole. It is a monotonic function and makes the spacetime asymptotically flat i.e., $\frac{S_f(r)}{r}$ tends to vanish for infinitely large values of the radial coordinate. Further, the shape function satisfies the flaring-out condition $\frac{S_f(r)-rS_f'(r)}{S_f(r)^2}>0$. This, at the throat, becomes $S_f'(r_0)<1$. Another significant function in describing the geometry of traversable wormhole is the proper radial distance function: 
 \begin{equation}\label{prd}
    L_f(r)=\pm \int_{r_0}^r \sqrt{\dfrac{r}{r-S_f(r)}}dr.
 \end{equation}
 The equation \eqref{prd} should be finite everywhere in the domain. Therefore $S_f(r)<r\;\forall\;r$ should be satisfied. The sign $\pm$ indicates respectively the upper and lower universes.

\par The gravitational interaction of the wormhole geometry with anisotropic matter distribution in $\mathpzc{f}(\mathcal{R},\mathscr{L}_m)$ gravity can be described using the field equations given by,
		\begin{widetext}
		\begin{eqnarray}
		    \label{fe1}4\mathpzc{f}_\mathcal{R}\dfrac{S_f'}{r^2}-(\mathpzc{f}-\mathpzc{f}_{\mathscr{L}_m}\mathscr{L}_m)=(2\rho+p_r+2p_\tau)\mathpzc{f}_{\mathscr{L}_m},\\
		    \label{fe2}	6\mathpzc{f}_\mathcal{R}''\left(1-\dfrac{S_f}{r} \right)+3\mathpzc{f}_\mathcal{R}'\left(\dfrac{S_f-rS_f'}{r^2} \right)+2\mathpzc{f}_\mathcal{R}\left(\dfrac{3S_f-rS_f'}{r^3} \right) 
				-(\mathpzc{f}-\mathpzc{f}_{\mathscr{L}_m}\mathscr{L}_m)=(-\rho-2p_r+2p_\tau)\mathpzc{f}_{\mathscr{L}_m}, \\
			\label{fe3}6\dfrac{\mathpzc{f}_\mathcal{R}''}{r}\left(1-\dfrac{S_f}{r} \right)-\mathpzc{f}_\mathcal{R}\left(\dfrac{3S_f-rS_f'}{r^3} \right)-(\mathpzc{f}-\mathpzc{f}_{\mathscr{L}_m}\mathscr{L}_m)=(-\rho+p_r-p_\tau)\mathpzc{f}_{\mathscr{L}_m}. 
		\end{eqnarray}
		\end{widetext}

\subsection{Energy Conditions}
\par  As a result of the Raychaudhuri equation, energy conditions govern the physical behavior of matter and energy in motion. One may examine the studies on the energy conditions in $\mathpzc{f}(\mathcal{R},\mathscr{L}_m)$ gravity in \cite{ecfrlm}. We will take into account the criteria for various energy conditions in order to assess the geodesic behavior. For the EMT \eqref{energymomentumtensor} with $\rho$, $p_r$ and $p_\tau$ respectively being energy density, radial pressure and tangential pressure, we have:
		\begin{enumerate}[label=$\circ$,leftmargin=*]
			\setlength{\itemsep}{4pt}
			\setlength{\parskip}{4pt}
			\setlength{\parsep}{4pt}
			\item \textit{Null Energy Conditions (NECs)}: $\rho+p_\tau\ge0$ and $\rho+p_r\ge0$.
			\item \textit{Weak Energy Conditions (WECs)}: $\rho\ge0\implies$  $\rho+p_\tau\ge0$ and $\rho+p_r\ge0$.
			\item \textit{Strong Energy Conditions (SECs)}: $\rho+p_j\ge0\implies$  $\rho+\sum_j p_j\ge0   \ \forall\ j$.
			\item \textit{Dominant Energy Conditions (DECs)}: $\rho\ge0\implies$ $\rho-|p_r|\ge0$ and $\rho-|p_\tau|\ge0$.
		\end{enumerate}
\section{Wormhole models in $\mathpzc{f}(\mathcal{R},\mathscr{L}_m)$ gravity}\label{III}

		\par In this section, we shall consider a viable wormhole model to study the characteristics of wormhole geometry. In particular, we suppose the non-linear form given by,    
				\begin{equation}\label{mod}
		    \mathpzc{f}(\mathcal{R},\mathscr{L}_m)=\dfrac{\mathcal{R}}{2}+\mathscr{L}_m^\alpha,
		\end{equation}
	    where $\alpha$ is a free parameter. For $\alpha=1$ the case reduces to GR. We presume that the matter Lagrangian density $\mathscr{L}_m$ depends on energy density $\rho$ i.e., $\mathscr{L}_m=\rho$ \cite{lmrho2,whfrlm1,lmrho1,lmrho3,lmrho4,lmrho5}. Now, comparing the equations \eqref{fe1} and \eqref{fe3} for $\mathpzc{f}(\mathcal{R},\mathscr{L}_m)$ model \eqref{mod} we can get the expressions for radial and tangential pressures as, 
		\begin{align}
		    \label{pr}p_r&=-\dfrac{\rho}{\alpha}\left[(\alpha-1)+\dfrac{S_f}{r^3\rho^\alpha}\right], \\
		    \label{pt}p_\tau&=\dfrac{r S_f'+S_f}{2\alpha r^3 \rho^{\alpha-1}} -\rho.
		\end{align}
	
      \subsection{Gaussian energy density:}

	   \par The equation \eqref{gauss} describes the energy density for Gaussian distribution. With the physical parameters $\rho$ \eqref{gauss}, $p_r$ \eqref{pr} and $p_\tau$ \eqref{pt} the field equation \eqref{fe2} reduces to,

        \begin{equation}
            \frac{S_f'(r)}{r}=8^{-\alpha } \pi ^{-\frac{3 \alpha }{2}} r \left(\frac{M e^{-\frac{r^2}{4 \theta }}}{\theta ^{3/2}}\right)^{\alpha }.
        \end{equation}
        \par On solving the above ordinary differential equation, the shape function of the wormhole with Gaussian distribution can be obtained. This is given by,
	 \begin{equation}\label{Asf}
	      \begin{split}
	         S_f(r)=\frac{2^{1-3 \alpha } \pi ^{-\frac{3 \alpha }{2}} \theta  \left[\sqrt{\pi\theta }\; e^{\frac{\alpha  r^2}{4 \theta }} \text{erf}\left(\frac{\sqrt{\alpha } r}{2 \sqrt{\theta }}\right)-\sqrt{\alpha } r\right] \left(\frac{M e^{-\frac{r^2}{4 \theta }}}{\theta ^{3/2}}\right)^{\alpha }}{\alpha ^{3/2}}+k,
	         \end{split}
	 \end{equation}
    where, $\text{erf} (z)=\frac{2}{\sqrt{\pi}}\int_0^z e^{-t^2} dt$ is the Gauss error function and $k$ is the integrating constant. Now, to obtain the particular solution, we find the value of $k$ by imposing the throat condition $S_f(r_0)=r_0$. Then we have,
    \begin{equation}\label{k1}
           \begin{split}
               k=\frac{\sqrt{\alpha }\; r_0 \left[\left(8 \pi ^{\frac{3}{2}}\right)^{\alpha} \alpha +2 \theta  \left(\dfrac{M e^{-\frac{r_0^2}{4 \theta }}}{\theta ^{\frac{3}{2}}}\right)^{\alpha }\right]-2 \sqrt{\pi }\; \theta ^{\frac{3}{2}}\; e^{\frac{\alpha  r_0^2}{4 \theta }}\; \text{erf}\left(\frac{\sqrt{\alpha } r_0}{2 \sqrt{\theta }}\right) \left(\dfrac{M e^{-\frac{r_0^2}{4 \theta }}}{\theta ^{\frac{3}{2}}}\right)^{\alpha }}{\left(8 \pi ^{\frac{3}{2}}\right)^{\alpha}\alpha ^{\frac{3}{2}}}.
           \end{split}
           \end{equation}
    \begin{figure}[!]
            \centering
            \subfloat[$S_f(r)$\label{fig:Asf1}]{\includegraphics[width=0.43\linewidth]{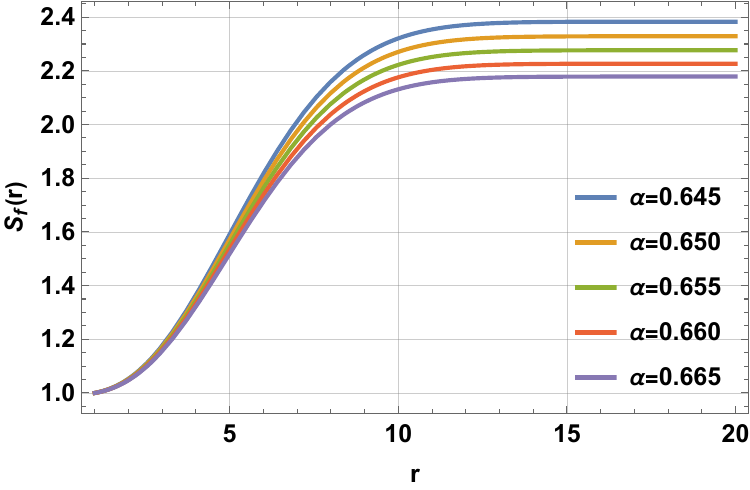}}
    	    \subfloat[$\frac{S_f(r)-rS_f'(r)}{S_f(r)^2}$\label{fig:Asf2a}]{\includegraphics[width=0.43\linewidth]{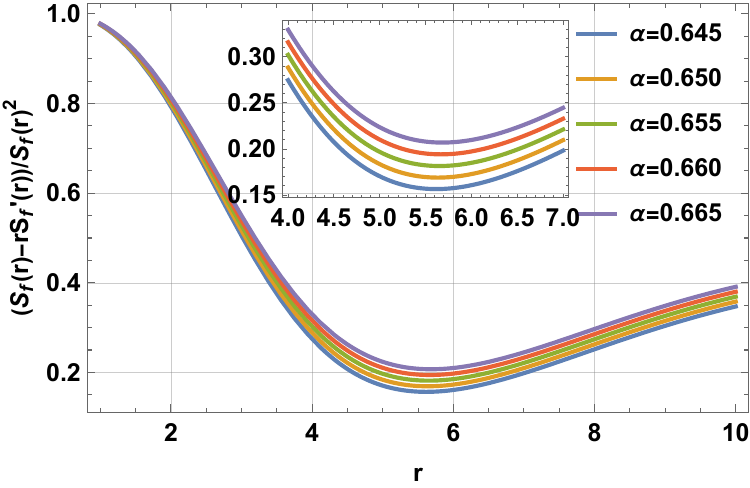}}\\
    	    \subfloat[$S_f'(r)$\label{fig:Asf2b}]{\includegraphics[width=0.43\linewidth]{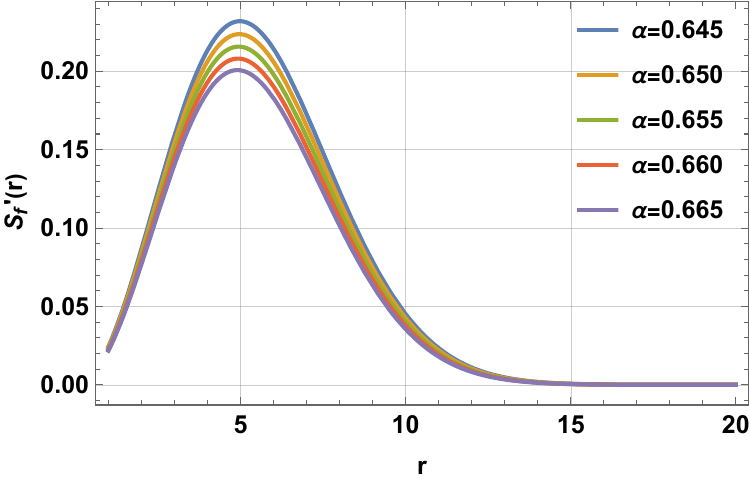}}
            \subfloat[$S_f(r)/r$\label{fig:Asf3}]{\includegraphics[width=0.43\linewidth]{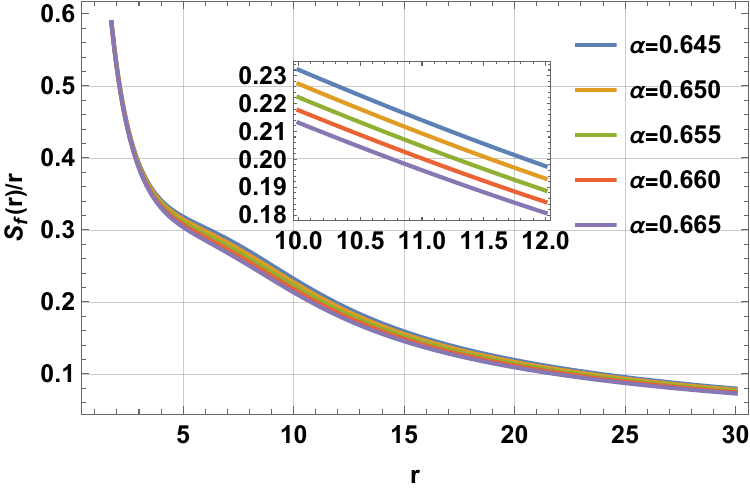}}
            \caption{Gaussian Distribution: The influence of model parameter $\alpha$ on the behavior of shape function $S_f$ with the total gravitational mass $M=1.2$ \textit{(mass)}, the square of minimal length $\theta=4$ \textit{(length$^{2}$)} and the throat radius $r_0=1$ \textit{(length)}.}
            \label{fig:Asf}
        \end{figure}

   In order to achieve the traversability of a wormhole the shape function should satisfy the flaring-out condition. For the present scenario, $S_f(r)$ satisfies $S_f'(r_0)<1$ at the throat if the following inequality holds:
       \begin{equation}\label{ineq1}
       \left(\dfrac{M\;e^{-\frac{1}{4\theta}}}{8\pi^{\frac{3}{2}}\theta^{\frac{3}{2}}}\right)^\alpha<\frac{1}{r_0^2}.
    \end{equation} 

    \begin{widetext}
	\begin{figure*}[t!]
	    \centering
	    \subfloat[Energy density $\rho$\label{fig:Arho}]{\includegraphics[width=0.39\linewidth]{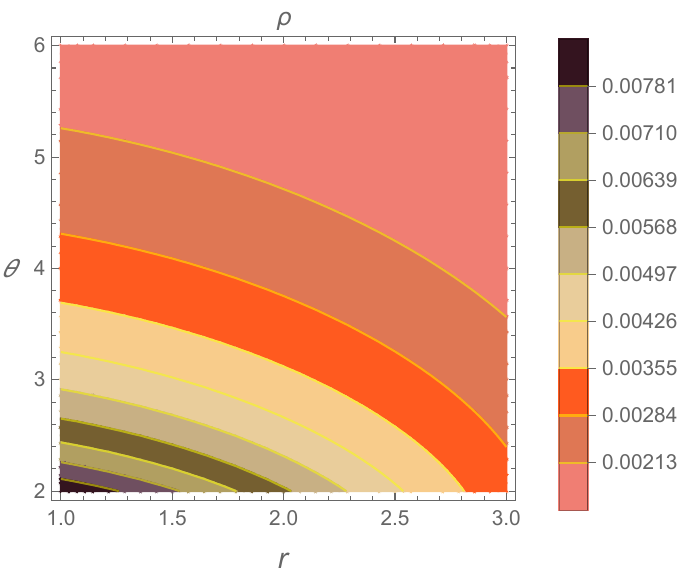}}
	    \subfloat[NEC $\rho+p_r$\label{fig:Ae1}]{\includegraphics[width=0.4\linewidth]{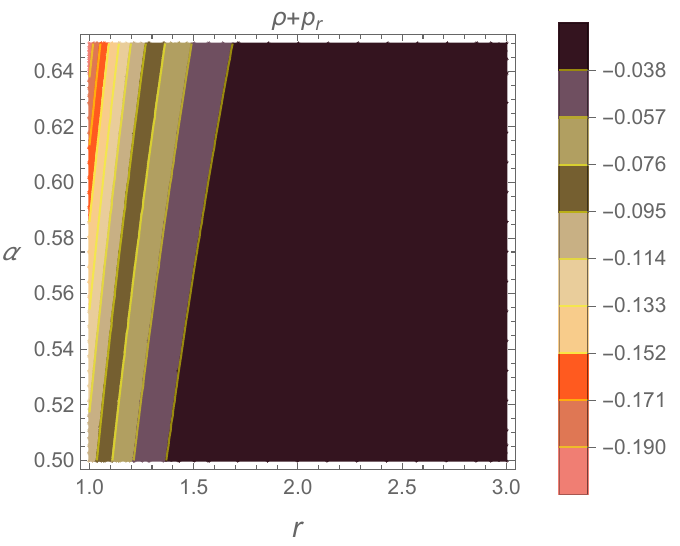}}\\
	    \subfloat[NEC $\rho+p_\tau$\label{fig:Ae2}]{\includegraphics[width=0.4\linewidth]{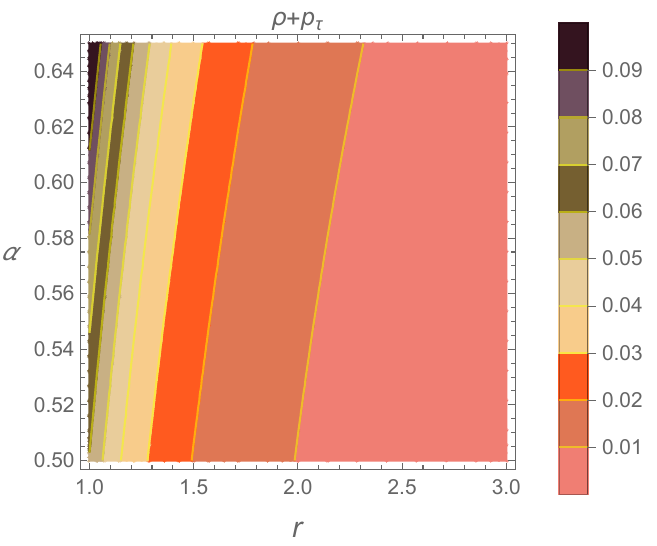}}
	    \subfloat[DEC $\rho-|p_r|$\label{fig:Ae3}]{\includegraphics[width=0.4\linewidth]{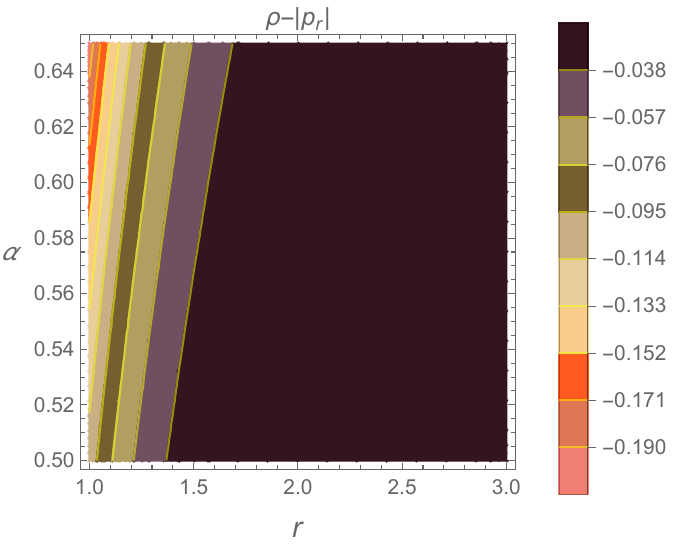}}\\
	    \subfloat[DEC $\rho-|p_\tau|$\label{fig:Ae4}]{\includegraphics[width=0.4\linewidth]{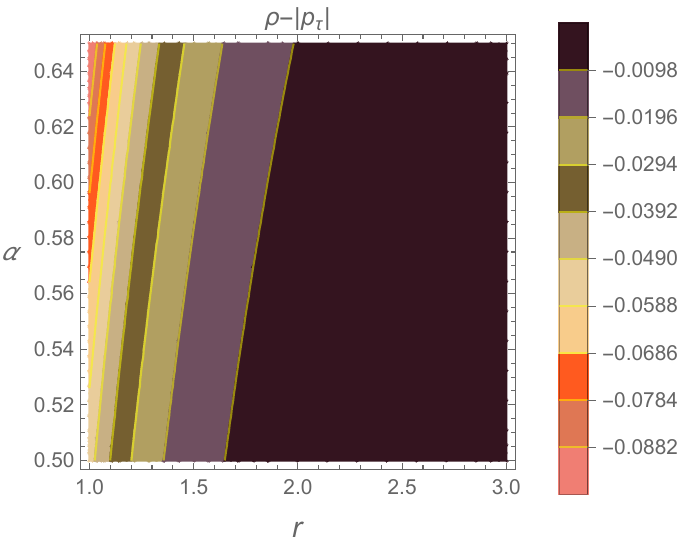}}
	    \subfloat[SEC $\rho+p_r+2p_\tau$\label{fig:Ae5}]{\includegraphics[width=0.4\linewidth]{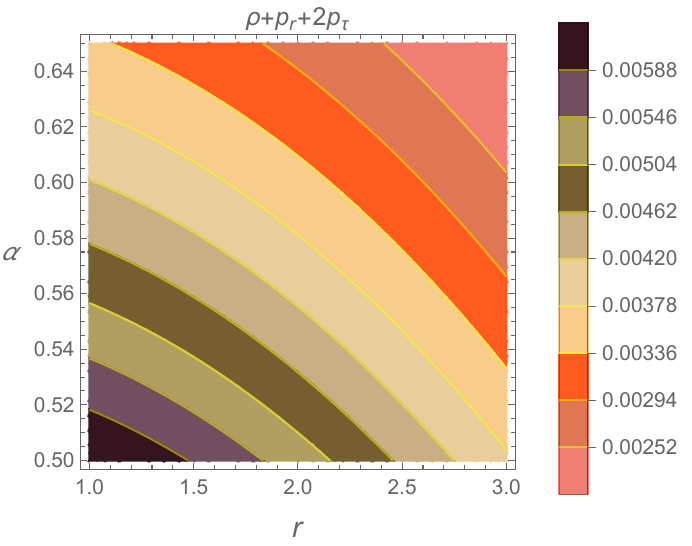}}
	    \caption{Gaussian Distribution: plot showing the profile of (a) energy density varying w.r.t radial coordinate $r$ and the square of minimal length $\theta$ \textit{(length$^{2}$)} with the total gravitational mass $M=1.2$ \textit{(mass)}, (b)-(f) different energy conditions varying w.r.t radial coordinate $r$ and model parameter $\alpha$ with the total gravitational mass $M=1.2$ \textit{(mass)}, the square of minimal length $\theta=4$ \textit{(length$^{2}$)} and the throat radius $r_0=1$ \textit{(length)}.}
	    \label{fig:Aec}
	\end{figure*}
    The above inequality is significant in determining the relation between $M$, $\theta$, $\alpha$, and $r_0$. For the GR scenario, with $r_0=1$ and $\theta=4$, we can get the constraining relation on the total mass $M$ as, $M<64\pi^{3/2} e^{1/16}$.  By taking the plot of shape function $S_f(r)$ with respect to $r$, we examined its behavior for $M=1.2, \theta=4$, and $r_0=1$. One can refer to TABLE$\;$\ref{tab:table0} for detailed analysis. For different values of $\alpha$, FIG.$\;$\ref{fig:Asf} shows the characteristics of $S_f(r)$. It is obvious from our choice of $k$ that $S_f(r)$ satisfies the throat condition. From FIG.$\;$\ref{fig:Asf1}, it can be seen that $S_f(r)>0$ is a monotonically increasing function. Moreover, $S_f(r)<r$, implying the finiteness of the proper radial distance function. Also, the flaring-out condition is satisfied [FIG.$\;$\ref{fig:Asf2a}, \ref{fig:Asf2b}]. The Lorentzian manifold becomes flat asymptotically as the value  $S_f(r)/r\to0$ for large $r$ and this can be interpreted from FIG.$\;$\ref{fig:Asf3}.

    Substituting equations \eqref{gauss}, \eqref{Asf}, \eqref{k1} in \eqref{pr} and \eqref{pt}, the pressure elements take the form,
    \begin{eqnarray}
         \begin{split}
            p_r=\frac{\left(\frac{M e^{-\frac{r^2}{4 \theta }}}{\theta ^{3/2}}\right)^{1-\alpha }}{8 \pi ^{3/2} \alpha ^{5/2} r^3} \left[-2 \sqrt{\pi } \theta ^{3/2} e^{\frac{\alpha  r^2}{4 \theta }} \text{erf}\left(\frac{\sqrt{\alpha } r}{2 \sqrt{\theta }}\right) \left(\frac{M e^{-\frac{r^2}{4 \theta }}}{\theta ^{3/2}}\right)^{\alpha }+2 \theta  \left(\sqrt{\pi } \sqrt{\theta } e^{\frac{\alpha  r_0^2}{4 \theta }} \text{erf}\left(\frac{\sqrt{\alpha } r_0}{2 \sqrt{\theta }}\right)-\sqrt{\alpha } r_0\right) \left(\frac{M e^{-\frac{r_0^2}{4 \theta }}}{\theta ^{3/2}}\right)^{\alpha } \right.\\\left.+\sqrt{\alpha } \left(-r \left((\alpha -1) \alpha  r^2-2 \theta \right) \left(\frac{M e^{-\frac{r^2}{4 \theta }}}{\theta ^{3/2}}\right)^{\alpha }-8^{\alpha } \pi ^{\frac{3 \alpha }{2}} \alpha  r_0\right)\right],
        \end{split}\\
        \begin{split}
            p_\tau=\frac{\left(\frac{M e^{-\frac{r^2}{4 \theta }}}{\theta ^{3/2}}\right)^{1-\alpha } }{16 \pi ^{3/2} \alpha ^{5/2} r^3} \left[2 \sqrt{\pi } \theta ^{3/2} e^{\frac{\alpha  r^2}{4 \theta }} \text{erf}\left(\frac{\sqrt{\alpha } r}{2 \sqrt{\theta }}\right) \left(\frac{M e^{-\frac{r^2}{4 \theta }}}{\theta ^{3/2}}\right)^{\alpha }+2 \theta  \left(\sqrt{\alpha } r_0-\sqrt{\pi } \sqrt{\theta } e^{\frac{\alpha  r_0^2}{4 \theta }} \text{erf}\left(\frac{\sqrt{\alpha } r_0}{2 \sqrt{\theta }}\right)\right) \left(\frac{M e^{-\frac{r_0^2}{4 \theta }}}{\theta ^{3/2}}\right)^{\alpha }\right.\\\left.+\sqrt{\alpha } \left(8^{\alpha } \pi ^{\frac{3 \alpha }{2}} \alpha  r_0-r \left(2 \theta +\alpha  (2 \alpha -1) r^2\right) \left(\frac{M e^{-\frac{r^2}{4 \theta }}}{\theta ^{3/2}}\right)^{\alpha }\right)\right].
        \end{split}
    \end{eqnarray}
       
    \par Furthermore, with the above pressure elements, we examined the energy conditions NEC, DEC, and SEC for Gaussian distribution [ref FIG.$\;$\ref{fig:Aec}]. In this case, NEC is not satisfied for radial pressure [FIG.$\;$\ref{fig:Ae1}] but for tangential pressure it holds [FIG.$\;$\ref{fig:Ae2}]. Also, both DECs are violated and SEC is satisfied [FIG.$\;$\ref{fig:Ae3},\ref{fig:Ae4},\ref{fig:Ae5}]. In addition, $\text{NEC}_{eff}\equiv \frac{rS_f'-S_f}{r^3}$ is violated as a consequence of satisfied flaring-out condition.

	 \begin{table*}[h!]
		\caption{Gaussian Distribution: Summarizing the nature of shape function $S_f$ with $M=1.2,\theta=4$ and $r_0=1$.}
		    \label{tab:table0}
		    \centering
		    \begin{tabular}{|c|c|c|}
		        \hline
		        \textit{Function }           & \textit{Result}  & \textit{Interpretation} \\
		        \hline
		        $S_f(r)$  & \makecell{ $0<S_f(r)<r\;\forall\;r>r_0$ and  $\; S_f(r_0)=r_0$ for $\alpha>0.41$}  & \makecell{Viable form of shape function\\ and throat condition is satisfied}\\
		        \hline
		        $\frac{S_f(r)-rS_f'(r)}{S_f(r)^2}$ & \makecell{$>0$, for $\alpha\ge0.58$ and $S_f'(r_0)<1$} & Flaring-out condition is satisfied \\
		       \hline
		       $\dfrac{S_f(r)}{r}$& \makecell{approaches to 0 for large value of $r$ and $\alpha>0$}& Asymptotic flatness condition is satisfied\\
		       \hline
			\end{tabular}
		\end{table*}
      \end{widetext}

    \paragraph{Wormhole Solutions:} In the context of Gaussian distribution, we have verified the criteria satisfied by the wormhole, such as finite redshift, throat condition, asymptotic condition, flaring-out condition, and the violation of effective null energy condition. Based on the constraining relation \eqref{ineq1}, we chose $M=1.2, \theta = 4$ and $r_0=1$ and analyzed the influence of model parameter $\alpha$ on the wormhole solution. For different values of $\alpha$ (say, $\alpha_i=0.645,0.650,0.655,09.660,0.656$), corresponding wormhole metrics read:
    \begin{equation}
        ds^2=e^{c}dt^2-\psi_i dr^2  - r^2\left(d\theta^2+\text{sin}^2\theta \,d\phi^2\right),
    \end{equation}
    where $c$ is some constant. For shape functions $S_{f_i}$ corresponding to $\alpha_i$, $\psi_i\equiv\left(1-\frac{S_{f_i}}{r}\right)^{-1}$, with

    \begin{align}
        \psi_1=&\frac{r}{\left(e^{-\frac{r^2}{16}}\right)^{0.645} \left(0.315227 r-1.39139 e^{0.0403125 r^2} \text{erf}(0.20078 r)\right)+r-0.99173},\\
        \psi_2=&\frac{r}{\left(e^{-\frac{r^2}{16}}\right)^{0.65} \left(0.304023 r-1.33676 e^{0.040625 r^2} \text{erf}(0.201556 r)\right)+r-0.991964},\\
        \psi_3=&\frac{r}{\left(e^{-\frac{r^2}{16}}\right)^{0.655} \left(0.293234 r-1.2844 e^{0.0409375 r^2} \text{erf}(0.20233 r)\right)+r-0.992191},\\
        \psi_4=&\frac{r}{\left(e^{-\frac{r^2}{16}}\right)^{0.66} \left(0.282845 r-1.23419 e^{0.04125 r^2} \text{erf}(0.203101 r)\right)+r-0.992411},\\
        \psi_5=&\frac{r}{\left(e^{-\frac{r^2}{16}}\right)^{0.665} \left(0.272839 r-1.18605 e^{0.0415625 r^2} \text{erf}(0.203869 r)\right)+r-0.992626}.
    \end{align}
    
    \subsection{Lorentzian energy density:}

    \par The non-commutative geometric distribution is an intrinsic aspect of a Lorentzian manifold \cite{intrinsic}. It is independent of the spacetime properties such as curvature. In this section, we study the scenario of the traversable wormhole with Lorentzian energy density distribution \eqref{lorentz}. Substituting \eqref{lorentz}, \eqref{pr} and \eqref{pt}, the field equation \eqref{fe3} becomes,

    \begin{equation}\label{ode2}
        \frac{S_f'(r)}{r}=\pi ^{-2 \alpha } r \left(\frac{\sqrt{\theta } M}{\left(\theta +r^2\right)^2}\right)^{\alpha }.
    \end{equation}

    \par The aforementioned equation is significant in determining the desired shape function. It is known that the shape function $S_f(r)$ at the throat should have a fixed point i.e., $S_f(r_0)=r_0$. Therefore, the ordinary differential equation \eqref{ode2} is an initial value problem. The particular solution of this equation is obtained as,
    
     \begin{equation}\label{Bsf}
	      \begin{split}
	         S_f(r)=\frac{r^3}{3\pi ^{2 \alpha }}   \left( M\theta ^{-\frac{3}{2}}\right)^{\alpha } \, _2F_1\left(\frac{3}{2},2\alpha ;\frac{5}{2};-\frac{r^2}{\theta }\right) +k,
	         \end{split}
	 \end{equation}
    where, ${}_2F_1(a,b;c;z)$ is the hypergeometric function and $k$ is the constant of integration given by,
    \begin{equation}
           \begin{split}
               k=r_0-\frac{r_0^3}{3\pi ^{2 \alpha }}   \left( M\theta ^{-\frac{3}{2}}\right)^{\alpha } \, _2F_1\left(\frac{3}{2},2\alpha ;\frac{5}{2} ;-\frac{r_0^2}{\theta }\right).
           \end{split}
           \end{equation}
   Additionally, we consider a constraining relation
   \begin{equation}\label{ineq2}
       \left(\dfrac{M\sqrt{\theta}} {\pi^2(1+\theta)^2}\right)^\alpha<\dfrac{1}{r_0^2}, 
   \end{equation}   
   in order to satisfy the flaring-out condition at the throat. For $\alpha=1$ we can retain the inequality for GR. With $r_0=1$ and $\theta=4$ the inequality \eqref{ineq2} reads, $M<25\pi^2/2$.

    \par
\begin{figure}[!]
        \centering
        \subfloat[$S_f(r)$\label{fig:Bsf1}]{\includegraphics[width=0.43\linewidth]{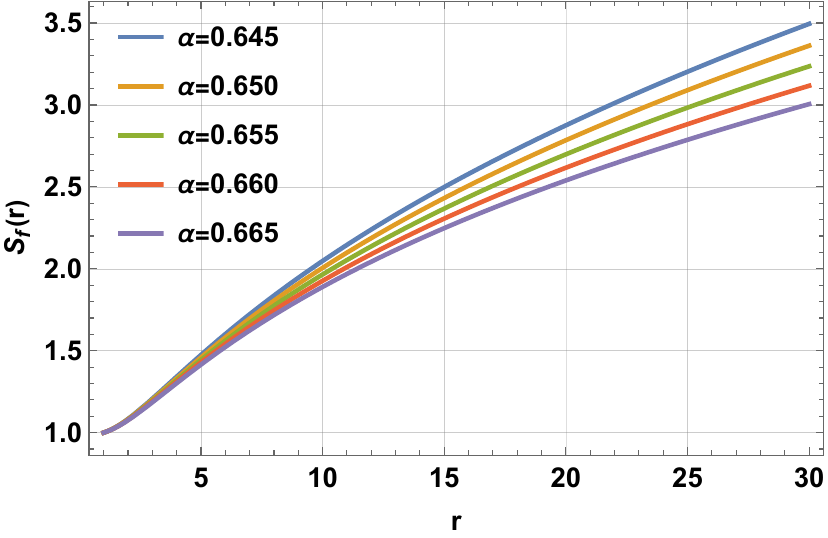}}
	    \subfloat[ $\frac{S_f(r)-rS_f'(r)}{S_f(r)^2}$\label{fig:Bsf2a}]{\includegraphics[width=0.44\linewidth]{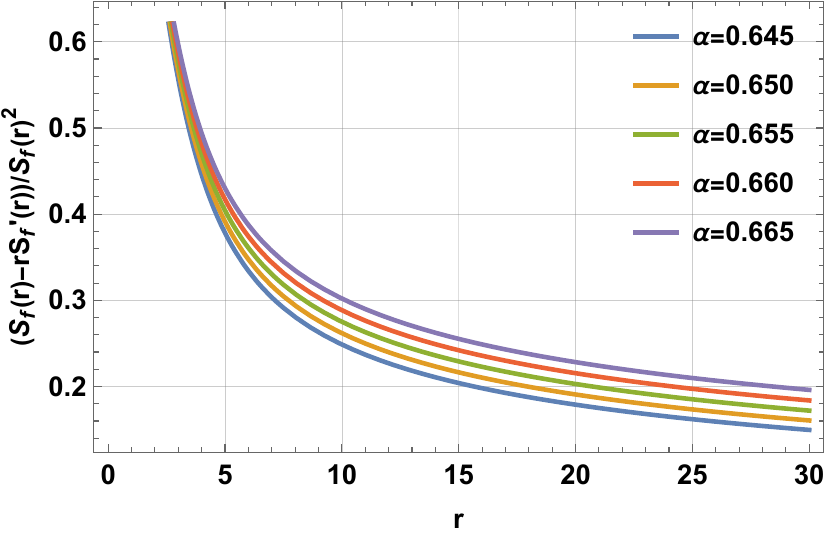}}\\
        \subfloat[$S_f'(r)$\label{fig:Bsf2b}]{\includegraphics[width=0.44\linewidth]{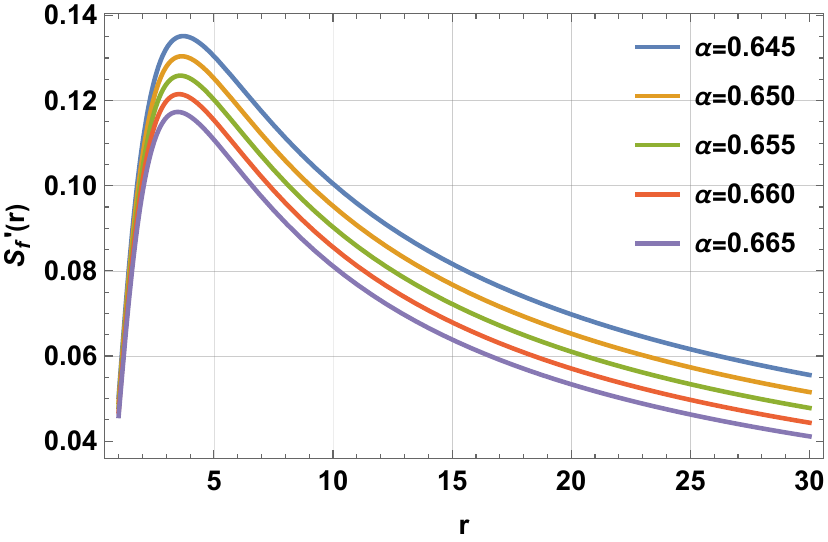}}
	    \subfloat[$S_f(r)/r$\label{fig:Bsf3}]{\includegraphics[width=0.43\linewidth]{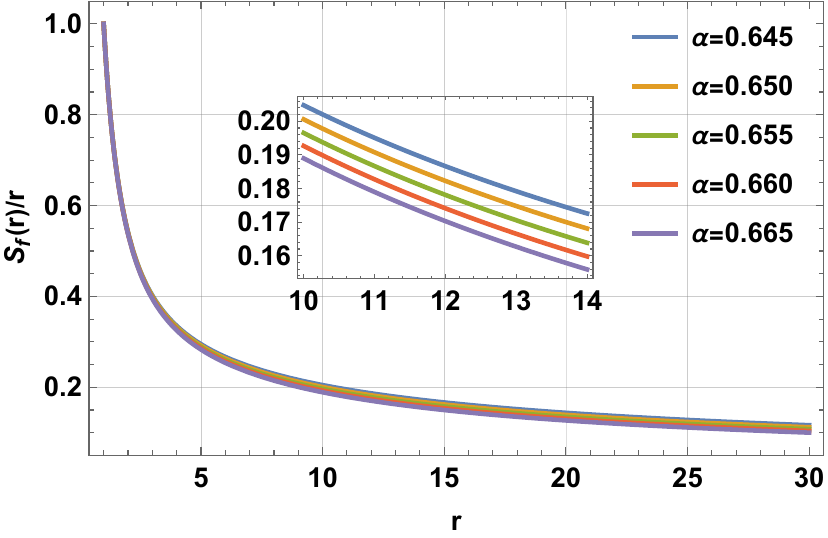}}
        \caption{Lorentzian Distribution: The influence of model parameter $\alpha$ on the behavior of shape function $S_f$ with the total gravitational mass $M=1.2$ \textit{(mass)}, the square of minimal length $\theta=4$ \textit{(length$^{2}$)} and the throat radius $r_0=1$ \textit{(length)}.}
        \label{fig:Bsf}
    \end{figure}

 Now we have to choose the range of values for the model parameter for which the obtained shape function satisfies all the necessary requirements. To this end, with the help of the plot of shape function versus radial coordinate, we studied the behavior of $S_f(r)$. The value of $\alpha$ is constrained to get the viable form of the shape function with Lorentz distribution [refer TABLE$\;$\ref{tab:table1}]. The effect of model parameter $\alpha$ on $S_f(r)$ for $M=1.2, \theta=4$ and $r_0=1$ is depicted in FIG.$\;$\ref{fig:Bsf}. It can be observed that $S_f(r)$ is a non-negative monotonically increasing function in the domain of radial coordinate $r$ [FIG.$\;$\ref{fig:Bsf1}] and satisfies the condition $S_f(r)<r$. Further, FIG.$\;$\ref{fig:Bsf2a}, \ref{fig:Bsf2b} reveals that the shape function obeys the flaring-out condition. For an infinitely large value of the radial coordinate $S_f(r)/r$ approaches to zero [FIG.$\;$\ref{fig:Bsf3}]. Thus, we can say that the shape function so obtained for the Lorentz distribution satisfies all the essential conditions.

    Further, with the shape function \eqref{Bsf} and energy density \eqref{lorentz}, the radial and tangential pressures can be rewritten as, 

    \begin{eqnarray}
        \begin{split}
            p_r=\frac{\left(\frac{\sqrt{\theta } M}{\left(\theta +r^2\right)^2}\right)^{1-\alpha }}{3 \pi ^2 \alpha  \theta  r^3} \left[-r^3 \left(\theta +r^2\right) \, _2F_1\left(1,\frac{5}{2}-2 \alpha ;\frac{5}{2};-\frac{r^2}{\theta }\right) \left(\frac{\sqrt{\theta } M}{\left(\theta +r^2\right)^2}\right)^{\alpha }-3 (\alpha -1) \theta  r^3 \left(\frac{\sqrt{\theta } M}{\left(\theta +r^2\right)^2}\right)^{\alpha }\right.\\\left.-3 \pi ^{2 \alpha } \theta  r_0+r_0^3 \left(\theta +r_0^2\right) \, _2F_1\left(1,\frac{5}{2}-2 \alpha ;\frac{5}{2};-\frac{r_0^2}{\theta }\right) \left(\frac{\sqrt{\theta } M}{\left(\theta +r_0^2\right)^2}\right)^{\alpha }\right].
        \end{split}\\
        \begin{split}
            p_\tau=\frac{\left(\frac{\sqrt{\theta } M}{\left(\theta +r^2\right)^2}\right)^{1-\alpha }}{6 \pi ^2 \alpha  \theta  r^3} \left[r^3 \left(\theta +r^2\right) \, _2F_1\left(1,\frac{5}{2}-2 \alpha ;\frac{5}{2};-\frac{r^2}{\theta }\right) \left(\frac{\sqrt{\theta } M}{\left(\theta +r^2\right)^2}\right)^{\alpha }-3 (2 \alpha -1) \theta  r^3 \left(\frac{\sqrt{\theta } M}{\left(\theta +r^2\right)^2}\right)^{\alpha }\right.\\\left.-r_0^3 \left(\theta +r_0^2\right) \, _2F_1\left(1,\frac{5}{2}-2 \alpha ;\frac{5}{2};-\frac{r_0^2}{\theta }\right) \left(\frac{\sqrt{\theta } M}{\left(\theta +r_0^2\right)^2}\right)^{\alpha }+3 \pi ^{2 \alpha } \theta  r_0\right].
        \end{split}
    \end{eqnarray}

    \par In addition, energy conditions interpret the characteristics of motion of energy and matter. Here, we studied the behavior of various energy conditions for Lorentz distribution with $M=1.2, \theta=4$, and $r_0=1$. The NEC is violated for radial pressure, supporting the requirement of the exotic fluid. Further, SEC and tangential NEC are obeyed. There is a violation of both the DECs. 

    \par In the next section, we shall analyze the physical aspects of the Gaussian and Lorentzian wormholes.
    
	\begin{figure*}[t!]
	    \centering
	    \subfloat[Energy density $\rho$\label{fig:Brho}]{\includegraphics[width=0.39\linewidth]{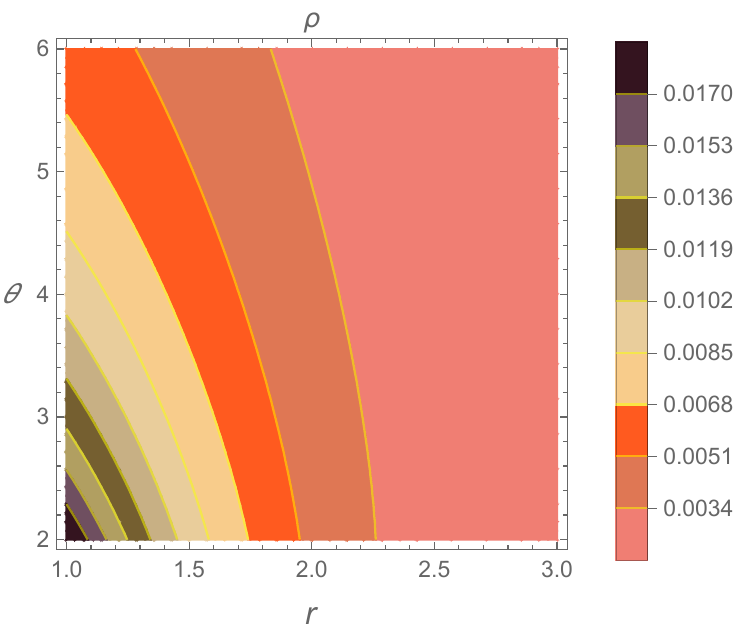}}
	    \subfloat[NEC $\rho+p_r$\label{fig:Be1}]{\includegraphics[width=0.4\linewidth]{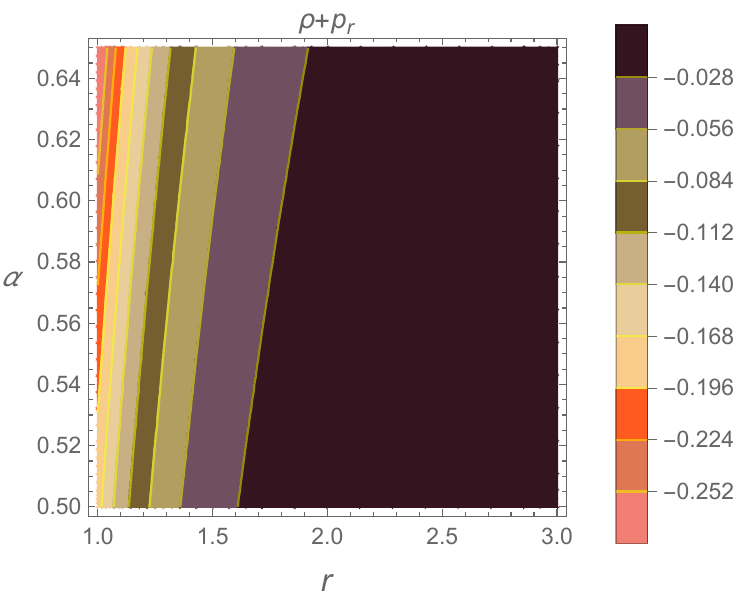}}\\
	    \subfloat[NEC $\rho+p_\tau$\label{fig:Be2}]{\includegraphics[width=0.4\linewidth]{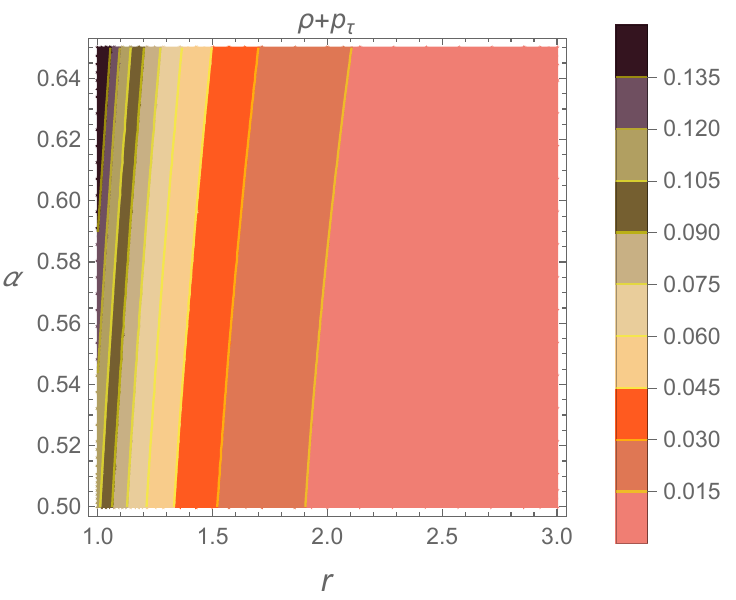}}
	    \subfloat[DEC $\rho-|p_r|$\label{fig:Be3}]{\includegraphics[width=0.4\linewidth]{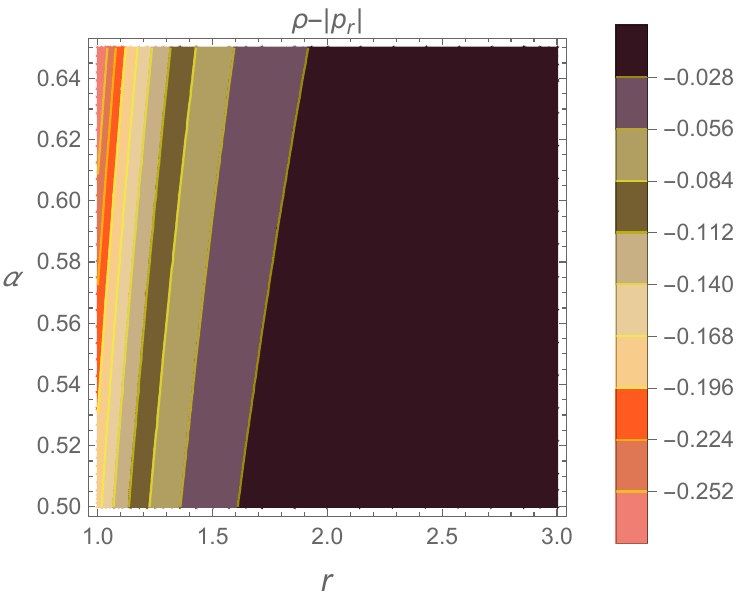}}\\
	    \subfloat[DEC $\rho-|p_\tau|$\label{fig:Be4}]{\includegraphics[width=0.4\linewidth]{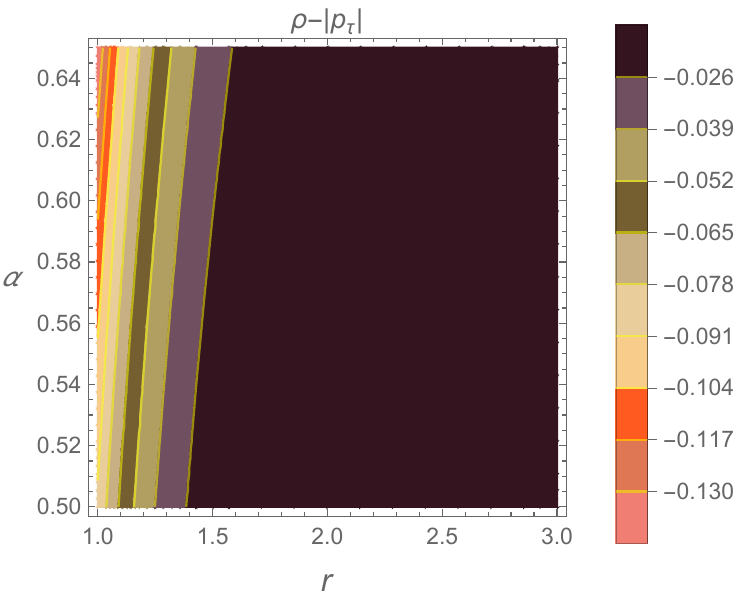}}
	    \subfloat[SEC $\rho+p_r+2p_\tau$\label{fig:Be5}]{\includegraphics[width=0.4\linewidth]{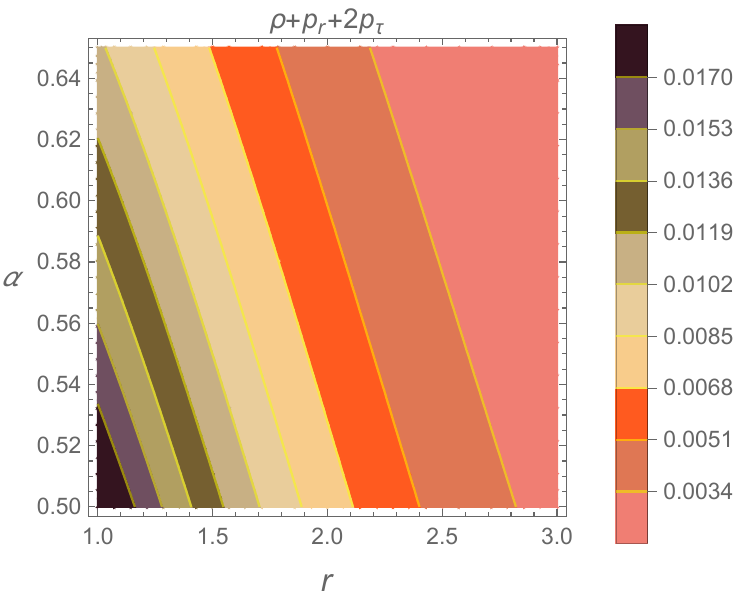}}
	    \caption{Lorentzian Distribution: plot showing the profile of (a) energy density varying w.r.t radial coordinate $r$ and the square of minimal length $\theta$ \textit{(length$^{2}$)} with the total gravitational mass $M=1.2$ \textit{(mass)}, (b)-(f) different energy conditions varying w.r.t radial coordinate $r$ and model parameter $\alpha$ with with the total gravitational mass $M=1.2$ \textit{(mass)}, the square of minimal length $\theta=4$ \textit{(length$^{2}$)} and the throat radius $r_0=1$ \textit{(length)}.}
	    \label{fig:Bec}
	\end{figure*}

 \begin{table*}[h!]
		\caption{Lorentzian Distribution: Summarizing the nature of shape function $S_f$ with $M=1.2,\theta=4$ and $r_0=1$.}
		    \label{tab:table1}
		    \centering
		    \begin{tabular}{|c|c|c|}
		        \hline
		        \textit{Function }           & \textit{Result}  & \textit{Interpretation} \\
		        \hline
		        $S_f(r)$  & \makecell{ $0<S_f(r)<r\;\forall\;r>r_0$ and  $\; S_f(r_0)=r_0$ for $\alpha>0.5$}  & \makecell{Viable form of shape function\\ and throat condition is satisfied}\\
		        \hline
		        $\frac{S_f(r)-rS_f'(r)}{S_f(r)^2}$ & \makecell{$>0$, for $\alpha>0.52$ and $S_f'(r_0)<1$} & Flaring-out condition is satisfied \\
		       \hline
		       $\dfrac{S_f(r)}{r}$& \makecell{approaches to 0 for large value of $r$, if $\alpha>0.5$}& Asymptotic flatness condition is satisfied\\
		       \hline
			\end{tabular}
		\end{table*}

  \paragraph{Wormhole solutions:} Within the framework of the Lorentzian distribution, we have examined the criteria that a traversable wormhole should satisfy. These include finite redshift, throat condition, asymptotic condition, flaring-out condition, as well as the violation of the effective null energy condition. By utilizing the constraining relation denoted by \eqref{ineq2}, we have selected specific values for the parameters $M=1.2$, $\theta=4$, and $r_0=1$, enabling us to investigate the impact of the model parameter $\alpha$ on the wormhole solution. For various values of $\alpha$ (i.e., $\alpha_i=0.645, 0.650, 0.655, 0.660, 0.656$), the corresponding metrics describing the wormhole are as follows:

  \begin{equation}
        ds^2=e^{c}dt^2-\psi_i dr^2  - r^2\left(d\theta^2+\text{sin}^2\theta \,d\phi^2\right),
    \end{equation}

    where $\psi_i$'s are given by,

    \begin{align}
        \psi_1=& -\frac{44.6546 r}{1. \left(r^2+4\right)^{1.29} \left(\frac{1}{\left(r^2+4\right)^2}\right)^{0.645} r^3 \, _2F_1\left(1.29,\frac{3}{2};\frac{5}{2};-\frac{r^2}{4}\right)-44.6546 r+43.8154},\\
        \psi_2=& -\frac{45.5992 r}{1. \left(r^2+4\right)^{1.3} \left(\frac{1}{\left(r^2+4\right)^2}\right)^{0.65} r^3 \, _2F_1\left(1.3,\frac{3}{2};\frac{5}{2};-\frac{r^2}{4}\right)-45.5992 r+44.7611},\\
        \psi_3=& -\frac{46.5637 r}{1. \left(r^2+4\right)^{1.31} \left(\frac{1}{\left(r^2+4\right)^2}\right)^{0.655} r^3 \, _2F_1\left(1.31,\frac{3}{2};\frac{5}{2};-\frac{r^2}{4}\right)-46.5637 r+45.7268}, \\
        \psi_4=&-\frac{47.5487 r}{1. \left(r^2+4\right)^{1.32} \left(\frac{1}{\left(r^2+4\right)^2}\right)^{0.66} r^3 \, _2F_1\left(1.32,\frac{3}{2};\frac{5}{2};-\frac{r^2}{4}\right)-47.5487 r+46.7129}, \\
        \psi_5=& -\frac{48.5545 r}{1. \left(r^2+4\right)^{1.33} \left(\frac{1}{\left(r^2+4\right)^2}\right)^{0.665} r^3 \, _2F_1\left(1.33,\frac{3}{2};\frac{5}{2};-\frac{r^2}{4}\right)-48.5545 r+47.7199}.
    \end{align}
\section{Equilibrium condition}\label{IV}
    In this section, we shall analyze the stability of Gaussian and Lorentzian wormhole models. For this purpose, we use the Tolman-Oppenheimer-Volkov (TOV) equation \cite{tov}:
        \begin{equation}\label{tov}
			p_r'+\dfrac{\varpi'}{2}(\rho+p_r)+\dfrac{2}{r}(p_r-p_\tau)=0,
		\end{equation}
		where primes ($'$) represent the derivative with respect to the radial coordinate $r$ and $\varpi=2R_f$. The aforesaid equation describes the equilibrium phase of a wormhole with gravitational $F_g$, hydro-static $F_h$, and anisotropic $F_a$ forces. These forces are defined by,

		\begin{align}
			F_g&=-R_f'(\rho+p_r),\\
			\label{fh}F_h&=-p_r',\\
			\label{fa}F_a&=\dfrac{2}{r}(p_\tau-p_r).
		\end{align}

\begin{figure}[!]
        \centering
        \subfloat[Equilibrium picture for Gaussian distribution\label{fig:Aeq}]{\includegraphics[width=0.48\linewidth]{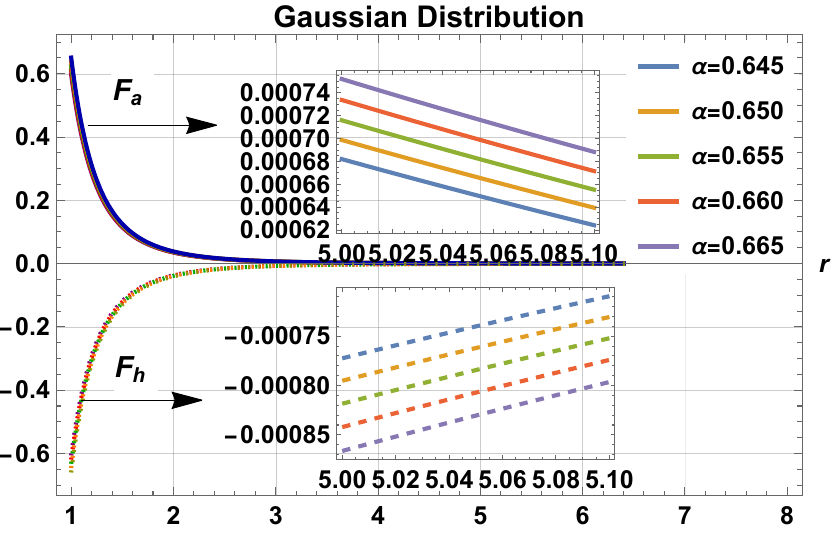}}
	    \subfloat[Equilibrium picture for Lorentzian distribution\label{fig:Beq}]{\includegraphics[width=0.48\linewidth]{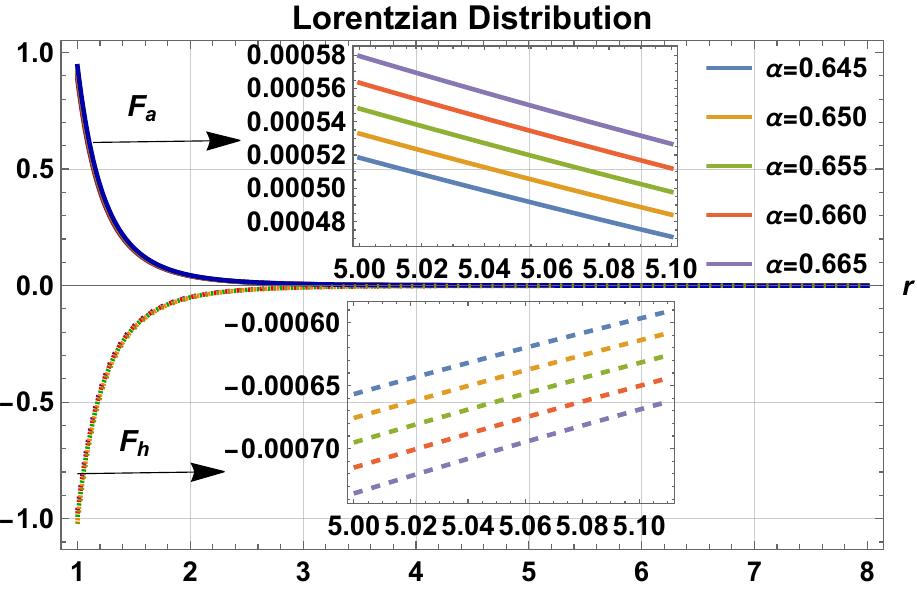}}
        \caption{The profile of hydro-static and anisotropic forces for different values of model parameter $\alpha$ with the total gravitational mass $M=1.2$ \textit{(mass)}, the square of minimal length $\theta=4$ \textit{(length$^{2}$)} and the throat radius $r_0=1$ \textit{(length)}.}
        \label{fig:eq}
    \end{figure}

        Thus, \eqref{tov} can be rewritten as, $F_g+F_h+F_a=0$. Since we have considered the tideless scenario, we have $R_f=0$ implying  $F_h+F_a=0$. FIG.$\;$\ref{fig:Aeq} and FIG.$\;$\ref{fig:Beq} illustrate the behavior of hydro-static and anisotropic forces for Gaussian distribution and Lorentzian distribution. From these plots, one can assess the influence of the model parameter on the equilibrium condition. It can be interpreted from FIG.$\;$\ref{fig:eq} that the nature of both hydro-static and gravitational forces are similar but opposite to one another. 

\section{Interpretation of wormhole}\label{V}
In this section, we shall discuss the physical aspects of the wormhole models.

\subsection{Average Pressure}
\par The average pressure $p$ can be described as,
$$p=\dfrac{1}{3}(p_r+2p_\tau).$$
For Gaussian distribution, the expression for average pressure reads,
    \begin{equation}
        p=\frac{(2-3 \alpha ) M e^{-\frac{r^2}{4 \theta }}}{24 \pi ^{3/2} \alpha  \theta ^{3/2}},
    \end{equation}
and for the Lorentzian distribution, it is given by,
    \begin{equation}
        p=-\frac{(3 \alpha -2) \sqrt{\theta } M}{3 \pi ^2 \alpha  \left(\theta +r^2\right)^2}.
    \end{equation}
    
\subsection{Speed of Sound}

\par The speed of sound parameter $v_s^2$ determines the stability of a wormhole. A wormhole is said to be stable if $0<v_s^2<1$ \cite{CZ8,sound1,sound2,sound3,sound4}. The speed of sound parameter is expressed as, 
$$v_s^2=\dfrac{dp}{d\rho}.$$
For both the Gaussian and Lorentzian distributions, this physical quantity is given by, 
\begin{equation}\label{sound}
    \dfrac{dp}{d\rho}=\frac{2}{3 \alpha }-1.
\end{equation}
One can note that $0\leq\dfrac{dp}{d\rho}<1$ is satisfied for $1/3\leq\alpha<2/3$. Therefore, non-commutative wormhole models are stable for $1/3\leq\alpha<2/3.$
\section{Results and Concluding Remarks}\label{VI}
\par Wormhole describes a path to connect two points of the universe. Recent research interests in this astrophysical entity are gaining importance due to its unique geometric feature. In the framework of modified theories of gravity, it is possible to construct physically viable wormhole structures. In our current study, we used an explicit coupling of matter with geometry which replaces the Ricci scalar in the Einstein-Hilbert action with an arbitrary function of the Ricci scalar and the matter Lagrangian. The geometry-matter coupling theory such as $\mathpzc{f}(\mathcal{R},\mathscr{L}_m)$ gravity can remarkably address the issue of exotic matter \cite{whfrlm1,gmc4}. On the other hand, non-commutative geometry with modified matter sources provides a mathematical approach to dealing with physical phenomena.
\begin{itemize}
    \item With the anisotropic matter distribution, we studied wormholes with zero tidal force. 
    \item Further, we presumed a non-linear $\mathpzc{f}(\mathcal{R},\mathscr{L}_m)$ model $\mathpzc{f}(\mathcal{R},\mathscr{L}_m)=\dfrac{\mathcal{R}}{2}+\mathscr{L}_m^\alpha$ with $\alpha$ being a free parameter. From \cite{frlm3,frlm4,frlm5}, we can see that the model is capable to explain the present scenario of the universe. 
    \item In the first case, we examined the wormhole scenario with Gaussian distribution. Here, we derived a shape function satisfying the throat condition. We verified the range of values of the model parameter $\alpha$ for which the obtained shape function obeys the flaring-out and asymptotic flatness conditions [ref TABLE$\;$\ref{tab:table0}].
    \item In the second case, we considered Lorentzian distribution to analyze wormhole properties. The obtained shape function obeys all the necessary conditions for a traversable wormhole. The parameter values of $\alpha$ for which the shape function fulfills the criteria are represented in the TABLE$\;$\ref{tab:table1}.
    \item We verified the flaring-out condition at the throat for both non-commutative geometries. The obtained inequalities $\left(\frac{M\;e^{-\frac{1}{4\theta}}}{8\pi^{\frac{3}{2}}\theta^{\frac{3}{2}}}\right)^\alpha<\frac{1}{r_0^2}$ and $ \left(\frac{M\sqrt{\theta}}          {\pi^2(1+\theta)^2}\right)^\alpha<\frac{1}{r_0^2}$ constrained the parameter values $M$, $r_0$, and $\theta$. In addition, with the aid of the speed of sound parameter, we checked the stability of the wormhole. It is inferred from equation \eqref{sound} that these non-commutative wormholes are stable for $1/3\le\alpha<2/3$. Further, the equilibrium condition is verified through the TOV equation. 
    \item In FIG.$\;$\ref{fig:Asf} and \ref{fig:Bsf}, we analyzed the impact of model parameter $\alpha$ on the behavior of shape functions. It is necessary to note that a minute variation in the value of $\alpha$ can impact the nature of shape functions. The nature of both shape functions is similar to that of the results obtained by Shamir and his collaborators \cite{shape}, in the context of exponential gravity coupled with the matter.
    \item In addition, throughout the manuscript, we analyzed the influence of this parameter $\alpha$ on the physical properties of the wormhole. Also, in all figures, we plotted the profile for those values of $\alpha$ for which we obtain a physically plausible wormhole solution. Further, in our analysis, we ignored those values of the model parameter leading to the negatively defined energy density. FIG.$\,$\ref{fig:Arho} and \ref{fig:Brho} show the density profile for both Gaussian and Lorentzian distribution.
    \item In both cases, there is a violation of the NEC, implying the existence of an exotic matter source. This agrees with the result obtained in GR and various modified theories with non-commutativity. In \cite{res},  for the linear $\mathpzc{f}(Q)$ model with charge, NEC is violated. Also, in \cite{ncg}, the result is obtained in the absence of charge. A similar instance can be seen in \cite{mat} with matter coupling in teleparallel gravity, which indicates the presence of exotic matter. In the context of Rastall gravity, Mustafa et al. \cite{res1} obtained the wormhole solutions violating NEC. 
\end{itemize}
\par In all, the letter has presented a stable viable wormhole model in the framework of $\mathpzc{f}(\mathcal{R},\mathscr{L}_m)$ gravity with non-commutative distributions. 



\section*{Data Availability Statement}
There are no new data associated with this article.

\begin{acknowledgments}
N.S.K. and V.V. acknowledge DST, New Delhi, India, for its financial support for research facilities under DST-FIST-2019. 
\end{acknowledgments}


\end{document}